\documentstyle[onecolumn]{mn}
\newcommand{\etal}{{et al.}~}

\newcommand{\f}{\frac}
\newcommand{\p}{\partial}
\newcommand{\ap}{\approx}
\newcommand{\Om}{\Omega}
\newcommand{\de}{\delta}

\newcommand{\al}{\alpha}
\newcommand{\fd}{\tilde{\delta}}

\newcommand{\fF}{\widetilde{F}}

\newcommand{\bfj}{{\bf j}}
\newcommand{\bfr}{{\bf r}}
\newcommand{\bfx}{{\bf x}}

\newcommand{\bfk}{{\bf k}}

\newcommand{\bfq}{{\bf q}}

\newcommand{\bfp}{{\bf p}}
\newcommand{\bfu}{{\bf u}}
\newcommand{\bfS}{{\bf S}}
\newcommand{\calC}{{\cal C}}
\newcommand{\calE}{{\cal E}}
\newcommand{\calL}{{\cal L}}
\newcommand{\calS}{{\cal S}}
\newcommand{\calH}{{\cal H}}
\newcommand{\bc}{\begin{center}}
\newcommand{\be}{\begin{equation}}
\newcommand{\ee}{\end{equation}}
\newcommand{\ec}{\end{center}}
\newcommand{\lan}{\langle}
\newcommand{\ran}{\rangle}

\title[Eulerian perturbation theory in non-flat universes]
{Eulerian perturbation theory in non-flat universes:
second-order approximation}

\author[P. Catelan, F. Lucchin, S. Matarrese and L. Moscardini]
{Paolo Catelan$^1$,
Francesco Lucchin$^2$,
Sabino Matarrese$^3$
and Lauro Moscardini$^2$ \\
$^1$ Department of Physics-Astrophysics, Nuclear Physics Laboratory,
Keble Road, Oxford OX1 3RH, UK \\
$^2$ Dipartimento di Astronomia, Universit\`a di Padova,
vicolo dell'Osservatorio 5, I-35122 Padova, Italy \\
$^3$ Dipartimento di Fisica {\it Galileo Galilei},
Universit\`a di Padova,
via Marzolo 8, I-35131 Padova, Italy\\}

\begin{document}

\maketitle

\begin{abstract}
The problem of solving perturbatively the equations describing the evolution of
self-gravitating collisionless matter in an expanding universe considerably
simplifies when directly formulated in terms of the gravitational and velocity
potentials: the problem can be solved {\it exactly}, rather than
approximately, even for cosmological models with arbitrary density
parameter $\Omega$. The Eulerian approach we present here allows to calculate
the higher-order moments of the initially Gaussian density and velocity
fields: in particular, we compute the gravitationally induced skewness of the
density and velocity-divergence fields for any value of $\Omega$, confirming
the extremely weak $\Omega$-dependence of the skewness
previously obtained via Lagrangian perturbation theory.
Our results show that the separability assumption of higher-order
Eulerian perturbative solutions is restricted to the Einstein-de Sitter case
only.

\end{abstract}


\section{Introduction}

Any model of large-scale structure formation based on the gravitational
instability has to deal with the problem of the nonlinear evolution of the
density and peculiar velocity fields. In the simplest of these scenarios, the
mass content of the universe is treated as a cold fluid of dust.
In spite of the simplicity of the dust fluid model, the analysis of the
nonlinear gravitational evolution is extremely involved,
the Poisson, continuity and Euler equations being highly nonlinear and
nonlocal beyond the linear regime, even in their Newtonian version
(Peebles 1980; see also Kofman \& Pogosyan 1994, and references therein).
In the attempt of addressing fundamental cosmological issues, a big effort
has been devoted to the theoretical treatment of nonlinear gravity, both in
the Eulerian and Lagrangian
formulation (e.g. Shandarin \& Zel'dovich 1989; Sahni \& Coles 1994).

The basic Eulerian approximation scheme is perturbation theory: the
density contrast $\de\equiv (\rho - \rho_b)/\rho_b$
($\rho$ being the density field and $\rho_b$ its mean background value)
and the peculiar velocity $\bfu$ are expanded about the
background solution $\de =0$, $\bfu = {\bf 0}$, namely $\de=\sum\de^{(n)}$
and $\bfu = \sum\bfu^{(n)}$ with $\de^{(n)}= O(\de^{(1)n})$ and
$\bfu^{(n)}= O(\bfu^{(1)n})$, where $\de^{(1)}$ and $\bfu^{(1)}$ correspond
to the linear solutions, then the differential equations for
any $\de^{(n)}$ and $\bfu^{(n)}$ are solved (Doroshkevich \& Zel'dovich 1975;
Peebles 1980; Fry 1984; Goroff \etal 1986).
Thus, for example, the question of whether power can be transferred
from large to small scales has been examined in the framework of the
second-order approximation by Juszkiewicz (1981), Vishniac (1983),
Juszkiewicz, Sonoda \& Barrow (1984), Coles (1990), Suto \& Sasaki (1991),
Makino, Sasaki \& Suto (1992), Jain \& Bertschinger (1994) and
Baugh \& Efstathiou (1994).
The weakly nonlinear deformation of the original Gaussian density
and velocity distributions in terms of the gravitationally induced
higher-order moments - and the corresponding correlation hierarchy -
is instead analysed by Peebles (1980), Fry (1984), Goroff \etal (1986),
Grinstein \etal (1987), Grinstein \& Wise (1987), Bernardeau (1992a,b),
Juszkiewicz, Bouchet \& Colombi (1993), Bernardeau (1994), Catelan \&
Moscardini (1994a,b), Juszkiewicz et al. (1994) and Lokas et al. (1994);
extensions to more general intrinsically non-Gaussian
models are discussed by Luo \& Schramm (1993) and Fry \& Scherrer (1994).
Some of these results and related issues have been also successfully
investigated in complementary N-body simulations by Hansel \etal (1985), Coles
\& Frenk (1991), Efstathiou \etal (1990), Efstathiou, Sutherland \& Maddox
(1990), Moscardini \etal (1991), Messina
\etal (1992), Weinberg \& Cole (1992), Bouchet \& Hernquist (1992), Lahav
\etal (1993), Lucchin \etal (1994), Kofman \etal (1994) and Melott, Pellman \&
Shandarin (1994).

Typically, with the exception of a few papers (Bernardeau 1992a,b; 1994),
this kind of investigation has been confined within the limits
of the Einstein-de Sitter cosmology, mainly because the condition $\Om=1$
enormously eases the solution of the dynamical equations
for $\de$ and $\bfu$: apart from theoretical expectations, there is however
no conclusive observational evidence that our universe is actually flat
(e.g. Peebles 1991; Coles \& Ellis 1994).

A powerful alternative approach is the {\it Lagrangian} description, where the
individual fluid elements are followed during their motion. The
most popular Lagrangian approach is the celebrated Zel'dovich
approximation (Zel'dovich 1970a,b), widely used in cosmology, showing also
extremely useful in reconstruction methods of initial conditions from velocity
data (e.g. Nusser \& Dekel 1992).

A Lagrangian perturbative approach, pioneered by Buchert (1989),
Moutarde \etal (1991) and Buchert (1992), is
presently object of thorough investigation. The key difference with respect to
the Eulerian approach is that one searches for solutions of
{\it perturbed trajectories} about the linear (initial) particle displacement
$\bfS^{(1)}:\;\bfS=\sum\bfS^{(n)}$ where $\bfS^{(n)} =  O (\bfS^{(1)})$,
while the continuity equation for the density field is exactly
satisfied.

Solutions up to the third-order Lagrangian approximation have been obtained,
in the case of an Einstein-de Sitter model by Buchert \& Ehlers (1993) and
Buchert (1994), and for more general Friedmann models by
Bouchet \etal (1992) and Catelan (1995).
Matarrese, Pantano \& Saez (1994a,b) developed a relativistic Lagrangian
treatment of the nonlinear dynamics of an irrotational collisionless fluid,
which reduces to the standard Newtonian approach on sub-horizon scales, but is
also suitable for the description of perturbations on super-horizon scales.

The higher accuracy of Lagrangian perturbative methods, as compared to other
currently studied approaches [such as the frozen flow (Matarrese \etal
1992) and frozen (or linear evolution of the) potential
(Brainerd, Scherrer \& Villumsen 1993; Bagla \& Padmanabhan 1994)
approximations, is
discussed by Munshi \& Starobinsky (1994), Bernardeau \etal (1994b) and
Munshi, Sahni \& Starobinsky (1994), still in the framework of an Einstein-de
Sitter cosmology. Comparisons with N-body simulations in the fully developed
nonlinear regime are given in Moutarde \etal (1991), Coles, Melott \&
Shandarin (1993), Melott \etal (1994) and Melott, Buchert \& Weiss (1995).

Finally, let us also mention some observational evidence that
gravitational instability in the quasi-linear regime operates on large scales.
Baumgart \& Fry (1991) claim that the observed bispectrum obeys the
hierarchical pattern; in Saunders \etal (1991), the distribution of
QDOT-{\it IRAS} galaxies out to $140 \,h^{-1}$Mpc ($h$ being
the Hubble constant in units of $100$ km sec$^{-1}$ Mpc$^{-1}$) is found
positively skewed and
inconsistent with the results of the standard cold dark matter model, even
after a 20 $h^{-1}$ Mpc smoothing (see also Coles \& Frenk 1991; Park 1991);
higher-order correlations
and/or moments in the galaxy distribution have been recently also measured by
Szapudi, Szalay \& Boschan (1992), Meiksin, Szapudi \& Szalay (1992), Bouchet
\etal (1993), Gazta\~naga (1992, 1994) and others, both in optically and
{\it IRAS} selected catalogues; higher-order effects are detected
on significantly
large scales by the counts-in-cells analysis of N-body simulations of
Baugh, Gazta\~naga \& Efstathiou (1994). The general observational
evidence appears to be quite consistent with gravitational instability
perturbation theory applied on random-phase initial conditions.

The problem we deal with in this paper is thus of quite general interest. We
present an Eulerian formalism which allows to solve in a rigorous
perturbative way the equations governing a pressureless self-gravitating
fluid in the expanding universe, with arbitrary density parameter $\Om$.
The problem notably simplifies when formulated directly in terms
of the peculiar gravitational and velocity potentials. In other words,
we look for perturbative solutions of the Bernoulli equation rather than
the Euler equation: the Bernoulli equation is the corner-stone for
several nonlinear approximation techniques (Nusser \etal 1991; Matarrese \etal
1992; Gramann 1993a; Lachi\`eze-Rey 1993; Brainerd, Scherrer \& Villumsen
1993; Bagla \& Padmanabhan 1994).

More in detail, we derive a particular form of the equations governing the
evolution of the gravitational potential $\varphi$ and the velocity potential
$\Phi$: the dynamics of the fields $\varphi$ and $\Phi$
is determined by the longitudinal component of the current $\de\bfu$ carried
by the fluctuation $\de$ with peculiar velocity $\bfu$. An advantage of this
treatment is that we can easily work out the higher-order velocity
contributions $\bfu^{(n)}$ in a manifestly irrotational form in Eulerian
configuration space [all the previous irrotational expressions
for $\bfu^{(n)}$ were obtained in Fourier space (see Goroff \etal 1986)]. The
importance of having explicit irrotational expressions for the velocity field
at any perturbative order lies in the fact that a collisionless
self-gravitating fluid cannot generate vorticity (before orbit mixing) and
that conservation of circulation is applicable well beyond the linear regime
(see the discussion in Bertschinger 1993).

Apart from the evident mathematical simplification, there are several more
reasons for working directly with the gravitational and velocity potentials.
Unlike the density fluctuation field, the gravitational potential
is required neither to have zero mean (since only its {\it gradients} are
physically meaningful), nor to satisfy the positive-mass constraint,
$\de\geq -1$; moreover, since it depends directly on the mass distribution,
it is bias-independent.
On the other hand, the velocity
potential has been recently recognized as a powerful
tool to test the cosmological power spectrum, becoming directly measurable
from observational data (see Heavens 1993; Baugh \& Efstathiou 1994).

There are two major previous attempts to solve the same problem we study in
this work. Martel \& Freudling (1991) applied Eulerian perturbation
theory, but in the attempt to get second-order solutions for the density
contrast and peculiar velocity fields in non-flat models, they
forced the intrinsic {\it non-separability}~ of the Eulerian dynamical
equations.
In Bouchet \etal (1992), the separability is also assumed, but
for the higher-order Lagrangian perturbative ansatz;
the separability hypothesis is also implicit in Gramann (1993a,b),
and Lachi\`eze-Rey (1993). Actually, the general validity of
the separability ansatz in Lagrangian perturbation theory has been recently
proved by Ehlers \& Buchert (1995, in preparation; see also Appendix B
in Catelan 1995).

As an application of our Eulerian formalism, we finally calculate the
density skewness $S_{3\de}$ induced by the
gravitational evolution, starting from Gaussian initial conditions,
as well as the skewness of the velocity-divergence, $S_{3\eta}$.

Before concluding this Introduction we would like to mention an
important difference between the Eulerian and Lagrangian perturbative
approaches, which is relevant for the present work.
Eulerian perturbation theory, at any order, cannot lead to shell-crossing
and consequent vorticity generation by multi-streaming, contrary to the
Lagrangian perturbation theory, and, what is most important, contrary to the
correct evolution which does lead to caustic formation.
This is because the perturbative Eulerian velocity field, in a given position
$\bfx$ and at time $t$, is obtained at any order by a set of (generally
nonlocal) quantities which are functions of the position $\bfx$ itself (and
time): no memory of the initial (Lagrangian) position $\bfq$ of the
particle which is passing there at that time can be present; in other terms,
the system does not contain particle inertia. Clear evidence of this
fact is provided by the frozen flow approximation (Matarrese \etal 1992),
which precisely exploits this property of the linear Eulerian velocity field
(but the same would be true at higher orders).
Therefore, provided the initial velocity field is sufficiently smooth,
which will be always the case after suitable low-pass filtering, no caustic
ever forms, as this would correspond to a discontinuity in the velocity field.
Of course, one might ask whether this is a real weakness of Eulerian
perturbation theory itself, compared to the Lagrangian one.
Certainly, it is a limitation if one is interested in using its results to
reconstruct the exact position of each particle after nonlinear evolution
(e.g. by extending the method of the frozen flow approximation to higher
orders). However, one is often interested in knowing only the
coarse-grained features of the density and velocity fields in physical space:
whether Eulerian theory at some perturbative order
can reproduce such features remains in our opinion an open question.
For the purpose of the present work the absence of multi-streaming
is a clear advantage, for it allows the use of the velocity potential at
any time during the nonlinear evolution.

The plan of the paper is as follows. In Section 2 the Hamilton-Jacobi
equation and the fluid evolution equations are derived starting from the
single particle Lagrangian in a generic Friedmann universe;
the Hamilton-Jacobi equation allows an alternative derivation of various
approximation schemes, such as the frozen flow and Zel'dovich algorithms,
which is sketched in Appendix A1, and shows an important analogy
with the Bernoulli equation for the velocity potential of the fluid. In
Section 3, the second-order solutions for the gravitational and velocity
potentials are worked out. These
results allow to calculate the gravitationally induced density and
velocity-divergence skewness and their dependence on the density parameter: in
Section 4 we compare these results with the corresponding ones obtained so far
by means of more complicated approximation techniques or applying
Lagrangian perturbation theory and transforming back to real space.
Conclusions are drawn in Section 5. Technical appendices are also
given.

\section{Dynamics of Self-Gravitating Collisionless Matter in Non-Flat
Universes}

We assume that, at the era of large-scale structure formation, the mass
content of the universe is in the form of a collisionless and irrotational
fluid with negligible velocity dispersion.
Let $\bfx$ be Eulerian comoving spatial coordinates; physical distances are
obtained by $\bfr=a(t)\bfx$, where $a(t)$ is the scale factor and $t$ the
cosmic time. For an Einstein-de Sitter universe (density parameter $\Om=1$ and
vanishing cosmological constant), $a(t) \propto t^{2/3}$. However, in what
follows, we shall consider more general non-flat Friedmann models.
In the general case a suitable time coordinate is the variable $D(t)$, in
that the dynamical equations assume very simple forms:
$D(t)$ is the growing mode of linear density fluctuations, whose
dependence on $t$ is in general quite complicated (e.g. Peebles
1980); with whole generality $D(0)=0$ at the initial time;
in the flat case, $D(t) \propto a(t) \propto t^{2/3}$.

In this section we rederive the dynamical equations for our pressureless
self-gravitating fluid directly from the single particle Lagrangian
in a non-flat model. The Hamilton-Jacobi formulation of the
problem is shown to be strictly related to the Bernoulli equation for the
velocity potential; in terms of it the Zel'dovich and
frozen flow approximations may be simply reformulated, as described in
Appendix A1.

\subsection{From the particles to the fluid description}

Let $\calL_t$ be the Lagrangian for the motion of a particle with mass $m$
(e.g. Peebles 1980)
\be
\calL_t(\bfx,\dot{\bfx}, t) =
\f{1}{2} m a^2 \,\dot{\bfx}^2 - m\, \phi(\bfx,t)\;,
\ee
where $\phi$ is the gravitational potential, related through the Poisson
equation to the mass fluctuations $\de$,
$\nabla^2 \phi = 4\pi G \, \rho_b \, a^2 \, \de$.
The action of the particle reads
$\calS = \int dt\, \calL_t = \int dD\,\calL_D$,
where the last equality is justified by the fact that $\calS$ is a scalar.
Here $dD = \dot{D}(t)dt$; in what follows, a dot indicates the
operator $d/dt$, and a prime $d/dD$. From now on, for simplicity,
we shall write $\calL \equiv\calL_D \equiv \dot{D}^{-1}\calL_t$, and
$\calL(\bfx,\bfx',D) = \f{1}{2} m a^2\dot{D}\,\bfx'^2 -
m\dot{D}^{-1}\,\phi(\bfx, D)$.
An alternative form of the Lagrangian $\calL$ may be obtained by
suitably rescaling the gravitational potential $\phi$ and the particle mass
$m$. Indeed, introducing the two quantities
$\varphi \equiv \left[2D/3e(\Om)a^2\dot{D}^2\right]\phi$
and $\mu(D) \equiv \left[a\dot{D}/\sqrt{D/e(\Om)}\,\right] m$,
after some algebra one gets the final expression
\be
\calL(\bfx, \bfx', D) = \f{a\mu}{2\sqrt{D/e(\Om)}}
\left\{ \f{D}{e(\Om)}\,\bfx'^2 - 3\,\varphi(\bfx, D) \right\}\;.
\ee
The function $e(\Om)$ is defined in terms of the logarithmic rate of growth of
mass fluctuations, $f(\Om) \equiv d\ln D/d\ln a$, through
$e(\Om)\equiv \Om/[f(\Om)]^2$ (Gramann 1993a,b). Note that
$e(\Om)$ depends weakly on $\Om$, since $e(\Om) \ap \Om^{-1/5}$, once
$f(\Om)\ap \Om^{3/5}$ is used (see Peebles 1980).
The Einstein-de Sitter case is easily recovered with the replacements
$D \to a$, $e \to 1$ and $\mu \to {2 \over 3} a_0^{3/2} t_0^{-1} m$
[here $t_0$ is an arbitrary reference time and $a_0=a(t_0)$].
In terms of the potential $\varphi$ the Poisson equation reduces to
\be
D\,\nabla^2\varphi(\bfx, D) = \delta(\bfx, D)\;.
\ee
Instead of using the velocity $\bfx'$ as independent variable,
one can alternatively introduce the conjugate momentum
$\bfp \equiv \f{\p\calL}{\p\bfx'} = a\sqrt{D/e(\Om)}\,\mu\,\bfx'$.
In such a case one adopts the Hamiltonian point of view, where
the Hamiltonian reads
\be
\calH(\bfx, \bfp, D) \equiv \bfp\cdot\bfx' - \calL(\bfx, \bfx', D) =
\f{\bfp^2}{2 a \mu \sqrt{D/e(\Om)}} +
\f{3}{2} \f{a\mu}{\sqrt{D/e(\Om)}}\,\varphi(\bfx, D)\;.
\ee
By varying the action $\calS$ with respect to $\bfx$
and $\bfx'$ one obtains the Euler-Lagrange equations
\be
\bfx'' + \f{3e(\Om)}{2D}(\bfx' + \nabla\varphi) = {\bf 0}\;.
\ee
The three second-order Euler-Lagrange equations above are equivalent to
the set of six first-order Hamilton equations, namely
\be
\bfx' = \f{\p\calH}{\p\bfp}= \f{\bfp}{a\mu\sqrt{D/e}}\;,
\;\;\;\;\;\;\;\;\;\;\;\;\;\;\;\;
\bfp'=-\f{\p\calH}{\p\bfx}=-\f{3}{2}\f{a\mu}{\sqrt{D/e}}\nabla\varphi\;.
\ee

Alternatively, one can use the Hamilton-Jacobi
formulation, according to which the momentum $\bfp$ is written in terms of
an action functional $\calS$, as $\bfp = \nabla\calS$, the latter
obeying the Hamilton-Jacobi equation
$\f{\p\calS}{\p D} + \calH(\bfx, \nabla\calS) = 0$.
It can be, however, useful to define a scaled action functional
$\Phi_{\calS}$ through
$\calS(\bfx, D) \equiv a\sqrt{D/e(\Om)}\,\mu\,\Phi_{\calS}(\bfx, D)$
so that the Hamilton-Jacobi equation takes the simple form, \
\be
\f{\p\Phi_{\calS}}{\p D} + \f{1}{2}(\nabla\Phi_{\calS})^2+\f{3e(\Om)}{2D}
(\Phi_{\calS} + \varphi) = 0\;.
\ee

So far we have only dealt with the dynamics of single particles. Our problem
is, however, the description of an infinite set of particles interacting only
through gravity. The standard description of this system
is based on the Vlasov equation (or collisionless Boltzmann equation),
which governs the evolution of the (comoving) one-particle distribution
function $f(\bfx,\bfp, D)d\bfx d\bfp$, namely the probability of finding one
particle in the infinitesimal phase-space element $d\bfx d\bfp$.

By the Liouville theorem the distribution function is conserved
along particle trajectories in phase-space, thus leading to the Vlasov
equation, which in our case reads
\be
\f{\p f}{\p D}+\f{\bfp}{a\mu\sqrt{D/e(\Om)}}\cdot\nabla f -
\f{3}{2}\f{a\mu}{\sqrt{D/e(\Om)}}\,\nabla\varphi\cdot\f{\p f}{\p \bfp}=0\;.
\ee
This equation, together with the definition of comoving mass density
\be
\varrho(\bfx, D) \equiv m ~n(\bfx, D) \equiv \rho_0 (1 + \de(\bfx, D))
\equiv m\int d\bfp\, f(\bfx,\bfp, D)
\ee
(here $n$ is the comoving particle number
density and $\rho_0$ the mass density at $t_0$)
and the Poisson equation (3), completes the description.
It is however useful to attempt a fluid description of the system,
which is achieved by
introducing the first moment of the distribution function, namely the local
streaming velocity, i.e. the mean velocity of the patch of fluid around $\bfx$
\be
\bfu(\bfx,D) = \f{1}{a\mu\sqrt{D/e(\Om)}}\;\f{\int d\bfp\,\bfp\,f(\bfx,\bfp,D)}
{\int d\bfp\,f(\bfx,\bfp,D)}\;.
\ee

{}From the Vlasov equation, we then obtain the continuity equation
\be
\f{\p\varrho}{\p D} + \nabla\cdot(\varrho\,\bfu) = 0 \;
\ee
and the momentum conservation equation
\be
\f{\p u_{\al}}{\p D} + u_{\beta} \p_{\beta}\, u_{\al}
+ \f{3 e(\Om)}{2D}\, \p_\al \varphi = - {1 \over \varrho} \p_\beta \bigl(
\varrho \,\pi_{\al\beta}\bigr) \;
\ee
(where $\p_{\al}\equiv \p/\p x_{\al}$; sum over repeated
indices is understood), having defined the velocity dispersion
tensor,
\be
\pi_{\al\beta}\equiv
\f{1}{a^2\mu^2 D/e(\Om)}\;\f{\int d\bfp\,p_{\al}\,p_{\beta}\,f(\bfx,\bfp,D)}
{\int d\bfp\,f(\bfx,\bfp,D)} - u_\al u_\beta\;.
\ee

For vanishing initial velocity dispersion and vorticity, and as long as
the system remains in the laminar regime we can write the distribution
function in the single-stream form
\be
f(\bfx,\bfp,D) = n(\bfx, D)\, \de_D
\Big(\bfp - a\sqrt{D/e(\Om)}\,\mu\,\nabla_{\bfx} \Phi_{\calS}(\bfx ,D)\Big)\;
\ee
[see also (Vergassola \etal 1994) for a related expression],
with $\de_D$ the Dirac delta function, which immediately leads to the
vanishing of the velocity dispersion tensor and to the validity of the fluid
picture. The resulting form of the momentum conservation equation is indeed
the Euler equation for a pressureless fluid in a general non-flat universe
(e.g. Shandarin 1994),
\be
{\p \bfu \over \p D} + (\bfu \cdot \nabla) \bfu
+ {3 e(\Om) \over 2 D} ( \bfu + \nabla \varphi) = 0\;.
\ee
Given the vanishing of the initial vorticity, Kelvin circulation theorem
implies that no vorticity will be generated at later times, except for
regions of multi-streaming (e.g. Doroshkevich 1973). In these conditions we
can define a velocity potential $\Phi$ via $\bfu = \nabla \Phi$. Replacing
this into the Euler equation one then finally gets the Bernoulli equation
\be
\f{\p\Phi}{\p D} + \f{1}{2}(\nabla\Phi)^2+\f{3e(\Om)}{2D}(\Phi +
\varphi) = 0\;,
\ee
which is formally {\em identical} to the scaled Hamilton-Jacobi
equation (7). One should, however, keep in mind that the action
$\Phi_{\calS}$ and the velocity potential $\Phi$ generally represent different
physical quantities. Moreover, while the validity of equation (7) has no
restrictions at all, the Bernoulli equation (16) is only valid provided
that i) one has vanishing initial vorticity and ii) vanishing initial velocity
dispersion, and iii) the system is in the single stream (or laminar)
regime, i.e. no orbit crossing has yet taken place.

\subsection{Dynamics of the gravitational and velocity potentials}

By expressing the density fluctuation $\de$ and velocity field $\bfu$ in terms
of the gravitational and velocity potential, respectively, we can recast the
continuity equation in the form (see also Kofman 1991)
\be
\nabla\cdot \left[\f{\p}{\p D}(D\nabla\varphi)+(1+D\nabla^2\varphi)\nabla\Phi
\right] = 0\;,
\ee
whose general solution reads
\be
\f{\p}{\p D}(D\nabla\varphi)+(1+D\nabla^2\varphi)\nabla\Phi=
\nabla\wedge{\bf T}\;.
\ee
A standard way to single out the irrotational part of the latter equation
is the following: according to Helmholtz's theorem (e.g. Morse \&
Feshbach 1953),
the last term in the l.h.s may be written as (up to an
irrelevant harmonic function)
\be
(D\nabla^2\varphi)\nabla\Phi=\de\bfu\equiv\nabla F +
\nabla\wedge{\bf T}\;,
\ee
in such a way that the solenoidal vector ${\bf T}$ in equation (18) is
{\it automatically} canceled out:
\be
\f{\p}{\p D}(D\nabla\varphi)+\nabla\Phi+\nabla F = 0 \;.
\ee
The function $F$ just introduced has to satisfy the equation
\be
\nabla^2F=\nabla\cdot(\de\bfu)= \de\nabla^2\Phi+
\nabla\de\cdot\nabla\Phi\;.
\ee

To summarize, one can write the set of equations for
the cosmological potentials as follows:
\be
\left\{
\begin{array}{l}
{\displaystyle \f{\p}{\p D}}\Big[D\varphi(\bfx,D)\Big]+\Phi(\bfx,D)+
F(\bfx,D)=0\;, \\ \\
{\displaystyle \f{\p}{\p D}}\Phi(\bfx,D) +
{\displaystyle \f{1}{2}}\Big[\nabla\Phi(\bfx,D)\Big]^2 +
{\displaystyle \f{3e(\Om)}{2D}}\Big[\Phi(\bfx,D)+\varphi(\bfx,D)\Big]=0\;,\\ \\
\nabla^2 F(\bfx,D)=D\nabla\cdot\left[\nabla^2\varphi(\bfx,D)\,
\nabla\Phi(\bfx,D)\right]\;.
\end{array}
\right.
\ee

Some remarks are appropriate. The auxiliary potential $F$ - which closes our
system of scalar equations - is related to the flux of matter originated by
the mass current $\bfj \equiv \de\bfu$
carried by the density fluctuation $\de$ moving with peculiar velocity $\bfu$:
only the longitudinal component $\bfj_{||}$ determines the dynamics
[a related quantity is introduced by Bertschinger \& Hamilton (1994),
although in a different context].
Furthermore, the function $F$ is, by construction, at least a second-order
quantity in the fields $\de$ and $\bfu$, because $\bfj$ is so.
This fact allows us to easily linearize the
equations: the first-order form of the equations (22) is
\be
\left\{
\begin{array}{l}
\varphi^{(1)} + \Phi^{(1)} = 0\;, \\ \\
{\displaystyle \f{\p \varphi^{(1)}}{\p D}} = 0\;.
\end{array}
\right.
\ee
so that the linear regime is described in terms of a single potential,
say $\varphi$ (see Kofman 1991). Note that the second of these
equations implies that the gravitational potential is time-independent, i.e.
$\varphi(\bfx,D) = \varphi^{(1)}(\bfx)$ and thus
$\Phi(\bfx,D)=\Phi^{(1)}(\bfx) = -\varphi^{(1)}(\bfx)$, in the linear regime.

Only a few exact solutions of the set of equations (22) are known,
mostly endowed with some symmetries, i.e. with restrictions in the initial
conditions. A possible alternative strategy is to seek perturbative
solutions: this program will be carried out in the next section.

\section{Eulerian Perturbation Theory: Second-Order Approximation}

Let us now solve the previous equations for the potentials according to
the following general $Eulerian$ perturbative ansatz:
\be
\varphi(\bfx,D)=
\varphi^{(1)}(\bfx)+\varphi^{(2)}(\bfx,D)+\cdots\equiv\varphi_1(\bfx)+
D\,\varphi_2(\bfx,D) +\cdots\;,
\ee
\be
\Phi(\bfx,D)=
\Phi^{(1)}(\bfx)+\Phi^{(2)}(\bfx,D)+\cdots
\equiv\Phi_1(\bfx)+D\,\Phi_2(\bfx,D)+\cdots\;,
\ee
explicitly up to second-order: of course, the expansion may be
continued to higher-order terms. The functions $\varphi^{(1)}(\bfx)$
and $\Phi^{(1)}(\bfx)$ are the linear potentials, as already discussed.
Note that in our notation, for instance,
$\de^{(1)}(\bfx,D)=D\,\nabla^2\varphi^{(1)}(\bfx)$, but
$\de_1(\bfx)=\nabla^2\varphi_1(\bfx)$, where
$\de^{(1)}(\bfx,D)\equiv D\,\de_1(\bfx)$; instead:
$\bfu^{(1)}(\bfx)\equiv\nabla\Phi^{(1)}(\bfx) =
\nabla\Phi_1(\bfx)\equiv\bfu_1(\bfx)$.
Similarly for
the higher-order terms: the expansions (24) and (25) obviously generate
analogous expansions for the corresponding fields $\de$ and $\bfu$.

We assume that the general ansatz in (24) and (25) actually catches {\it all}
the physical information contained in perturbation theory: in this sense
we stress that, in particular, the second-order solutions
$\varphi^{(2)}(\bfx,D)$ and $\Phi^{(2)}(\bfx,D)$ are {\it non-separable} in
the variables $\bfx$ and $D$. On the contrary, what is usually done is to
assume that all the temporal information contained e.g. in
$\varphi^{(2)}(\bfx,D)$ can be factored out in the form
$\varphi^{(2)}(\bfx, D)=D\,\varphi_2(\bfx)$: this is not allowed
{\it a priori},
it applies in the case of the Einstein-de Sitter universe, but not in a
generic Friedmann model. To be convinced
of the previous statement it is enough to suppose, {\it ab absurdo}, that
separability actually holds, then show that such an ansatz is
{\it not} a solution of the fundamental equations; we give
this proof in Appendix A2.

In terms of the perturbed quantities, our equations become
\be
\left\{
\begin{array}{l}
{\displaystyle D\f{\p}{\p D}}\varphi_2(\bfx,D)+2\,\varphi_2(\bfx,D)+
\Phi_2(\bfx,D)+F_2(\bfx)=0\;, \\ \\
{\displaystyle D\f{\p}{\p D}}\Phi_2(\bfx,D) +
\left[1+{\displaystyle \f{3e(\Om)}{2}}\right]\,\Phi_2(\bfx,D)+
{\displaystyle \f{3e(\Om)}{2D}}\varphi_2(\bfx,D) +
{\displaystyle \f{1}{2}}\Big[\nabla\Phi_1(\bfx)\Big]^2 =0\;,\\ \\
\nabla^2 F_2(\bfx)=\nabla\cdot\left[\nabla^2\varphi_1(\bfx)
\,\nabla\Phi_1(\bfx)\right]=\nabla \cdot \Big[\de_1(\bfx)\,\bfu_1(\bfx)\Big]\;,
\end{array}
\right.
\ee
where the dependence on the variables $\bfx$ and $D$ is explicitly
shown and, after noting that $F(\bfx,D)$ is actually a second-order
quantity, we have defined the scaled function
$F_2(\bfx)\equiv D^{-1}F^{(2)}(\bfx,D)$; the solution of the last equation
above can be immediately obtained as a convolution in Fourier space,
$\fF_2(\bfk) \equiv \int d\bfx\, F_2(\bfx)\, {\rm e}^{\,i\,\bfk\cdot\bfx}\,$:
\be
\fF_2(\bfk)=\f{1}{k^2}\int\f{d\bfk_1 d\bfk_2}{(2\pi)^3}
\,\de_D(\bfk_1+\bfk_2-\bfk)\,\left[1+\f{\bfk_1\cdot\bfk_2}{k_2^2}\right]\,
k_1^2\tilde{\varphi}_1(\bfk_1)\,k_2^2\tilde{\varphi}_1(\bfk_2)\;,
\ee
where e.g. $k=|\bfk|\,$: $F_2$
is determined by the initial condition on $\varphi_1$, while the kernel
$1+\bfk_1\cdot\bfk_2/k^2_2$ describes the effects of the nonlinear evolution.

The differential equations in (26) are coupled in the fields
$\varphi_2(\bfx,D)$ and $\Phi_2(\bfx,D)$; to decouple them, we
differentiate the first once, to get
\be
D\f{\p^2\varphi_2}{\p D^2} + 3\,\f{\p\varphi_2}{\p D} + \Phi_2 = 0\;.
\ee
By substituting into the second equation of the set (26) we obtain
\be
D^2\f{\p^2\varphi_2}{\p D^2} + \left[4+\f{3e(\Om)}{2}\right]D
\f{\p\varphi_2}{\p D}+
\left[2+\f{3e(\Om)}{2}\right]\varphi_2= \f{1}{2}(\nabla\varphi_1)^2
-\left[1+\f{3e(\Om)}{2}\right]F_2\;,
\ee
which turns out to be an equation for the single potential $\varphi_2$.
Now it should be noticed that, while $\varphi_2(\bfx,D)$ is not separable
in space and time, the sum
\be
\psi(\bfx, D) \equiv \varphi_2(\bfx,D) + F_2(\bfx)
\ee
is exactly separable; in fact, after substitution in equation (29), we find
\be
D^2\f{\p^2\psi}{\p D^2} + \left[4+\f{3e(\Om)}{2}\right]D\f{\p\psi}{\p D} +
\left[2+\f{3e(\Om)}{2}\right]\psi= C(\bfx)\;,
\ee
where
\be
C(\bfx)\equiv\f{1}{2}\Big[\nabla\varphi_1(\bfx)\Big]^2 +F_2(\bfx) \;.
\ee
Defining at this point the function
\be
A(\Om)\equiv 2+\f{3}{2}\,e(\Om)=A(D)\;,
\ee
we can summarize our results in terms of the following set of equations,
\be
\left\{
\begin{array}{l}
\left[{\displaystyle D^2\f{\p^2}{\p D^2}+\Big(2+A(D)\Big)D\f{\p}{\p D}}+
A(D)\right]
\psi(\bfx,D)=C(\bfx)\;, \\ \\
\varphi_2(\bfx,D)=\psi(\bfx,D)-F_2(\bfx)\;,\\ \\
\Phi_2(\bfx,D)={\displaystyle -D\f{\p}{\p D}}\varphi_2(\bfx,D)
-2\,\varphi_2(\bfx,D) - F_2(\bfx)\;,
\end{array}
\right.
\ee
which exactly defines the second-order potentials $\varphi_2(\bfx, D)$ and
$\Phi_2(\bfx, D)$. In particular, since $F_2(\bfx)$ is
determined by the initial conditions, the last equation of the set (34)
gives the
velocity potential $\Phi_2(\bfx,D)$ once $\varphi_2(\bfx, D)$
is known: the main task is to solve the first of the equations
in (34); we address this problem in the next subsection.

\subsection{Second-order perturbative solutions}

We have previously shown that, although the second-order potential
$\varphi_2(\bfx,D)$ is non-separable, the function
$\varphi_2(\bfx,D)+F_2(\bfx)$ is so: since the r.h.s. of the first equation in
(34) does not depend on $D$, we can assume (up to an additive
constant, which can be always set to zero)
\be
\psi(\bfx,D)\equiv B(D)\,C(\bfx)\;,
\ee
where the function $B(D)$ is a function of the only variable $D$. Therefore,
the second-order potential $\varphi_2(\bfx,D)$ takes the form
\be
\varphi_2(\bfx,D)= - \Big[1-B(D)\Big]F_2(\bfx)+
\f{1}{2}B(D)\Big[\nabla\varphi_1(\bfx)\Big]^2\;.
\ee
The second-order velocity potential then reads
\be
\Phi_2(\bfx,D) = \Big[1-2B(D)-DB'(D)\Big]F_2(\bfx)-
\f{1}{2}\Big[2B(D)+DB'(D)\Big]\Big[\nabla\varphi_1(\bfx)\Big]^2\;.
\ee
The last relation shows that the velocity $\bfu$
may be indeed written as the gradient of a scalar function in
$\{\bfx\}$-space:
\be
\bfu(\bfx,D) = \nabla\Big[\Phi^{(1)}+\Big(1-2B(D)-DB'(D)\Big)F^{(2)}(\bfx, D)-
\f{1}{2}D\Big(2B(D)+DB'(D)\Big)\bfu^{(1)2}\Big]\;,
\ee
explicitly up to second-order; this expression of the
velocity field is thus manifestly irrotational in physical space
[see also Catelan \& Moscardini (1994b); for its Fourier components
see Goroff \etal (1986)].

We can now furthermore derive the
final expressions of the second-order corrections
$\de^{(2)}(\bfx, D)\equiv D^2\,\de_2(\bfx, D)$ and
$\nabla\cdot\bfu^{(2)}(\bfx, D)\equiv D\,\nabla\cdot\bfu_2(\bfx, D)\,$:
\be
\de^{(2)}(\bfx,D)=\Big[1-B(D)\Big]\de^{(1)}(\bfx, D)^2-
D\left(\bfu^{(1)}\cdot\nabla\right)\de^{(1)}(\bfx, D) +
D^2 B(D)\sum_{\al\beta}\left(\p_{\al}u^{(1)\beta}\right)^2\;,
\ee
and
\begin{eqnarray}
-D\,\nabla\cdot\bfu^{(2)}(\bfx,D)=
\Big[1-2B(D)\!\!&-&\!\!DB'(D)\Big]\de^{(1)}(\bfx, D)^2 -
D\left(\bfu^{(1)}\cdot\nabla\right)\de^{(1)}(\bfx, D) \nonumber \\
\!\!&+&\!\!
D^2\Big[2B(D)+DB'(D)\Big]\sum_{\al\beta}\left(\p_{\al}u^{(1)\beta}\right)^2\;.
\end{eqnarray}
Combining these solutions, we can also derive
a relation between the density contrast $\de$ and the divergence of the
velocity field $\nabla\cdot\bfu$, namely
\be
-D\,\nabla\cdot\bfu = \de - 2\,D^2\Big[DB(D)\Big]'\,\mu_2(\bfu_1)\;,
\ee
where $\mu_2(\bfu_1)$ is the second-order invariant of the initial
deformation tensor $\p_{\al}\p_\beta \Phi_1(\bfx)$, which is in general
different from zero in three-dimensional systems:
$2\mu_2(\bfu_1)\equiv(\nabla\cdot\bfu_1)^2 -
\sum_{\al\beta}\left(\p_{\al}u_1^{\beta}(\bfx)\right)^2 =
2\,(\lambda_1\lambda_2+\lambda_1\lambda_3+\lambda_2\lambda_3)$.
The quantities $\lambda_h, h=1,2,3$, are the eingenvalues of
$\p_{\al}u_1^{\beta}(\bfx)$. The relation (41), which
quantifies the second-order deviation of $\nabla\cdot\bfu$ from $\de$,
may be compared with similar relations in the literature (e.g.
Giavalisco \etal 1993; Mancinelli
\etal 1993; Mancinelli \& Yahil 1994). The importance
of the shear term $\sum_{\al\beta}\left(\p_{\al}u_1^{\beta}(\bfx)\right)^2$
in the density-velocity-divergence relation, even in the case of laminar
flow, is also discussed by Mancinelli \& Yahil (1994).

The corresponding expressions of $\de^{(2)}$ and $\bfu^{(2)}$ for the flat
limit case (see Peebles 1980, where a different
gravitational potential is used; Goroff \etal 1986, where a different
normalization of the velocity field is chosen; also Appendix A3 for an
independent derivation) read:
\be
\de^{(2)}(\bfx,a)=\f{5}{7}\,\de^{(1)}(\bfx, a)^2-
a\left(\bfu^{(1)}\cdot\nabla\right)\de^{(1)}(\bfx, a) +
\f{2}{7}\,a^2 \sum_{\al\beta}\left(\p_{\al}u^{(1)\beta}\right)^2\;,
\ee
and
\be
-a\,\nabla\cdot\bfu^{(2)}(\bfx,a)= \f{3}{7}\,\de^{(1)}(\bfx, a)^2-
a\left(\bfu^{(1)}\cdot\nabla\right)\de^{(1)}(\bfx, a) +
\f{4}{7}\,a^2\sum_{\al\beta}\left(\p_{\al}u^{(1)\beta}\right)^2\;,
\ee
where, in the Einstein-de Sitter model, $\de^{(2)}$ scales like $a^2$
and $\bfu^{(2)}$ like $a$, in that they are separable in space and
time, $\de^{(2)}(\bfx, a)=a^2\,\de_2(\bfx)$ and
$\bfu^{(2)}(\bfx, a)=a\,\bfu_2(\bfx)$.

The function $B(D)$, defined in (35), allows to specify the
(second-order) growth of the potentials $\varphi(\bfx,D)$ and
$\Phi(\bfx,D)$: it is a solution of the differential equation
\be
\left\{D^2\f{d^2}{dD^2}+\Big[2+A(D)\Big]D\f{d}{dD}+A(D)\right\}
B(D)=1\;.
\ee
The initial condition (at the beginning: $D=0$) for the function
$B(D)$ may be easily found comparing equations (39) and (42):
$B(D=0)=2/7$.
Comparing instead equations (40) and (43), we get a relation for the
derivative of $B(D)$ at the initial time, namely $DB'(D)|_{D=0}=0$.

Since the dependence of $A(D)$ on $D$ is definitively non-trivial, we
transform to a new variable $\tau$ which, e.g. in the case $\Om < 1$, is
defined by (see e.g. Peebles 1980)
\be
\tau \equiv \f{1}{\Om} -1\;,
\ee
where $0 \leq \tau < \infty$: $\tau=0$ corresponds to the Big Bang
(where one asymptotically approaches the Einstein-de Sitter model). The
previous differential equation can now be written as
\be
\left\{
J(\tau)^2\f{d^2}{d\tau^2} + \Big[1+J'(\tau)+A(\tau)\Big]\,
J(\tau)\,\f{d}{d\tau}+
A(\tau)\right\}\,B(\tau) = 1\;,
\ee
and the initial conditions keep the same form.
Here and in what follows, a prime denotes differentiation with respect to
$\tau$. To maintain the notation concise
we have defined
\be
J(\tau) \equiv \f{D(\tau)}{D'(\tau)} = \tau\,f(\tau)^{-1}\;,
\ee
and $J(\tau) \to \tau$ in the limit $\tau \to 0$.
The growing mode $D(\tau)$ reads (e.g. Peebles 1980):
\be
D(\tau)= 1 + \f{3}{\tau} + \f{3\sqrt{1+\tau}}{\tau^{3/2}}\,L(\tau)\;;
\ee
with $L(\tau)\equiv{\rm ln}\Big(\sqrt{1+\tau}-\sqrt{\tau}\Big)$;
$D(\tau) \to 2\tau/5$ in the limit $\tau\to 0$.
The function $A(\tau)$ is now simply given by
\be
A(\tau)=2+\f{3}{2}\Big[(1+\tau)\,f(\tau)^2\Big]^{-1}\;,
\ee
where finally
\be
f(\tau) = -\f{3}{2}\left[
\f{3\sqrt{\tau(\tau+1)} + (3+2\tau)L(\tau)}
{(3+\tau)\sqrt{\tau(\tau+1)}+(3+3\tau)L(\tau)}
\right]\;
\ee
and $f(0) = 1$.

The analytical solution of Eq.(46) for the open case, with growing mode initial
conditions, $B(\tau)|_{\tau=0}=2/7$ and $\tau B'(\tau)|_{\tau=0}=0$, can be
guessed by noting that there exists a close relation between our Eq.(46) and
Eq.(7) in Bouchet \etal (1992), such that
$B=\f{1}{2}\left[1+\f{E(\tau)}{D(\tau)^2}\right]$, where $E(\tau)$
is also reported in eq.(70) below. Indeed, the function
\be
B(\tau) = \f{1}{2} -\f{1}{4D(\tau)^2}-\f{9}{4\tau D(\tau)^2}
\left\{1+\sqrt{\f{1+\tau}{\tau}}\,L(\tau)+\f{1}{2}
\left[\sqrt{\f{1+\tau}{\tau}}+\f{L(\tau)}{\tau}
\right]^2
\right\}\;,
\ee
solves the differential equation (46) with the appropriate initial conditions.
This result also shows the equivalence between the Eulerian perturbation
theory, which is in general characterized by non-separable perturbative
solutions, and the Lagrangian description, whose higher-order
solutions are intrinsically separable
(see the discussion in Appendix B of Catelan 1995 for the second-order case;
the separability of higher-order Lagrangian modes has been recently
demonstrated by Ehlers \& Buchert 1995, in preparation; Buchert, private
communication).

Incidentally we give here also
the corresponding expressions directly in terms of the variable $\Om$, which
can alternatively be chosen as time variable: in the order,
\be
J(\Om) = -\Om(1-\Om)\,f(\Om)^{-1}\;,
\ee
\be
D(\Om) = \f{1}{(1-\Om)^{3/2}}\left[(1+2\Om)\sqrt{1-\Om}+3\Om\, L(\Om)
\right]\;,
\ee
\be
f(\Om) = -\f{3\Om}{2}
\left[\f{3\sqrt{1-\Om}+(2+\Om)\,L(\Om)}{(1+2\Om)\sqrt{1-\Om}+3\Om\, L(\Om)}
\right]\;,
\ee
where $L(\Om)\equiv{\rm ln}\,\Big[(1-\sqrt{1-\Om}\,)/\sqrt{\Om}\,\Big]$ and
$L(1)=0$. There are two approximations of the function $f(\Om)$ usually
found in the literature: $f(\Om)\ap \Om^{3/5}$ (Peebles 1980) and
$f(\Om)\ap \Om^{4/7}$ (Fry 1985). Also, we give all the formulae for
the closed model ($\Om>1$) in Appendix A4.

\section{Applications: the Gravitational Skewness}

The previous results may be used to compute the gravitationally
induced skewness of the density fluctuation, $\de$, and divergence of the
velocity field, $\eta \equiv - D\,\nabla\cdot\bfu$. We
assume that the primordial
potential $\varphi$ is Gaussian distributed; then, for the
density contrast the first non-zero contribution to the skewness is
\be
\lan\de^3\ran = 3 \lan\de^{(1)2}\de^{(2)}\ran + O(\de^{(1)6})\;.
\ee
It is well known that this kind of calculations is better performed
in Fourier space, namely we need to compute the
term $\lan\fd^{(1)}(\bfk_1,D)\,\fd^{(1)}(\bfk_2,D)\,\fd^{(2)}(\bfk_3,D)\ran$.
{}From equation (39), one finally gets
\be
\fd_2(\bfk,D) = \int \f{d\bfk_1 d\bfk_2}{(2\pi)^3}\,
\de_D(\bfk_1+\bfk_2-\bfk)\,J_s^{(2)}(\bfk_1,\bfk_2;D)\,\fd_1(\bfk_1)\,
\fd_1(\bfk_2)\;,
\ee
where the symmetrized kernel $J_s^{(2)}$ is defined by the relation
\be
J_s^{(2)}(\bfk_1,\bfk_2;D) \equiv
\Big[1-B(D)\Big] + \f{1}{2}\left(\f{k_1}{k_2}+\f{k_2}{k_1}\right)\,
\f{\bfk_1\cdot\bfk_2}{k_1\,k_2} +
B(D)\left(\f{\bfk_1\cdot\bfk_2}{k_1\,k_2}\right)^2\;.
\ee
The gravitational skewness of the unfiltered density field may thus be
written in the form
\be
\lan\de^3\ran = 6\,\int\f{d\bfk d\bfk'}{(2\pi)^6}\,J_s^{(2)}(\bfk,\bfk';D)\,
P(k)\,P(k')\;,
\ee
where $P(k)$ is the primordial power spectrum of the density field.
Performing the integration, we obtain the following expression
for the density skewness parameter $S_{3\de}$:
\be
S_{3\de}(\tau) \equiv \f{\lan\de^3\ran}{\lan\de^2\ran^2} =
\f{3\lan\de^{(1)2}\de^{(2)}\ran}{\lan\de^{(1)2}\ran^2} = 6 - 4B(\tau)\;,
\ee
having emphasized the dependence on, e.g., the variable $\tau$. Note that,
at the beginning $B(0)=2/7$, and therefore $S_3(0)=34/7$, which is the
well-known value of the unfiltered density skewness in the Einstein-de
Sitter model (Peebles 1980).

Similar calculations can be done for the velocity-divergence field $\eta$. It
turns out that
\be
\tilde{\eta}_2(\bfk,D) = \int \f{d\bfk_1 d\bfk_2}{(2\pi)^3}\,
\de_D(\bfk_1+\bfk_2-\bfk)\,K_s^{(2)}(\bfk_1,\bfk_2;D)\,\fd_1(\bfk_1)\,
\fd_1(\bfk_2)\;,
\ee
where the kernel $K_s^{(2)}$ is defined by
\be
K_s^{(2)}(\bfk_1,\bfk_2;D) \equiv
\Big[1-2B(D)-DB'(D)\Big] + \f{1}{2}\left(\f{k_1}{k_2}+\f{k_2}{k_1}\right)\,
\f{\bfk_1\cdot\bfk_2}{k_1\,k_2} +
\Big[2B(D)+DB'(D)\Big]\left(\f{\bfk_1\cdot\bfk_2}{k_1\,k_2}\right)^2\;.
\ee
Then
\be
\lan\eta^3\ran = 6\,\int\f{d\bfk d\bfk'}{(2\pi)^6}\,K_s^{(2)}(\bfk,\bfk';D)\,
P(k)\,P(k')\;,
\ee
from which
\be
S_{3\eta}(\tau) \equiv
\f{\lan\eta^3\ran}{\lan\eta^2\ran^2} = 6 - 8B(\tau)-4J(\tau)B'(\tau)\;.
\ee

Note that, for the Einstein-de Sitter model, we recover the
standard result $S_{3\eta}(0)=26/7$ (see Bernardeau 1994).

\subsection{Previous determinations of the skewness}

Both the results (59) and (63) can be directly compared with
the corresponding results in the literature.

\subsubsection{Martel \& Freudling (1991)}

In Martel \& Freudling (1991), the second-order Eulerian equation
for the density field is reduced to a separable version of it, by replacing
$\Om^{1.2} \to \Om$ in one of the terms.
Martel and Freudling call this ``almost--separability" of the
perturbation equation for the density field, and the consequent
approximation ``MF approximation" (see also Martel 1995).
In practice, this corresponds to assume that the expansion of the
density contrast is of the factorized form
\be
\de(\bfx,\tau) =
D(\tau)\de_1(\bfx)\Big[1+\kappa(\tau)\calC(\bfx)+\cdots\Big]\;,
\ee
up to the second-order term. The function $\calC(\bfx)$ is given by the
expression
\be
\calC(\bfx) = \f{5}{2}\de_1(\bfx)-
\f{7}{8\pi\de_1(\bfx)}\nabla\de_1(\bfx)\cdot\nabla\chi(\bfx) +
\f{1}{16\pi^2\de_1(\bfx)}\left(\p_{\al}\p_{\beta}\chi(\bfx)\right)^2\;,
\ee
where $\chi\equiv \int d\bfx'\,\de_1(\bfx')/|\bfx'-\bfx|$; furthermore,
the function $\kappa(\tau)$ is a solution of the differential equation
\be
\tau^2\kappa''(\tau)+2\left(\Om^{3/5}+1-\f{\Om}{4}\right)\tau\kappa' =
(\tau+1)^{-1}D(\tau)\;,
\ee
with $\kappa(0)=\kappa'(0)=0$ and $\kappa(\tau)\to 2D(\tau)/7$ in
the limit $\tau\to 0$. It turns out that the density skewness is given
by the expression
\be
S_{3\de}(\tau)= 17\,\f{\kappa(\tau)}{D(\tau)}\;.
\ee
In the mentioned references no explicit estimate of $S_{3\eta}$ is
reported. \\

\subsubsection{Bouchet \etal (1992)}

The dynamical equations for the
displacement $\bfS(\bfq,\tau)$ from the Lagrangian (initial) positions
$\bfq$ of the fluid elements are solved in Bouchet \etal (1992) according to
the perturbative expansion
\be
\bfS(\bfq,\tau)=D(\tau)\,\bfS^{(1)}(\bfq)+E(\tau)\,\bfS^{(2)}(\bfq)+ \cdots\;.
\ee
up to the second-order correction [see also Catelan (1995) for the
notation]. After (perturbatively) inverting the relation
$\bfx = \bfq + S(\bfq,\tau)$,
to get the Eulerian location of the mass elements, one obtains the
following expression for the second-order density contrast $\de^{(2)}$:
\be
\de^{(2)}(\bfx, \tau) =
\f{1}{2}\Big[1-E/D^2\Big]\de^{(1)}(\bfx, \tau)^2 -
D\left(\bfu^{(1)}\cdot\nabla\right)\de^{(1)}(\bfx, \tau) +
\f{D^2}{2}\Big[1+E/D^2\Big]\sum_{\al\beta}\left(\p_{\al}u^{(1)\beta}\right)^2\;,
\ee
where the function $E(\tau)$ is given by the relation
\be
E(\tau) = -\f{1}{2}-\f{9}{2\tau}
\left\{1+\sqrt{\f{1+\tau}{\tau}}\,L(\tau)+\f{1}{2}
\left[\sqrt{\f{1+\tau}{\tau}}+\f{L(\tau)}{\tau}
\right]^2
\right\}\;,
\ee
which, in the limit $\tau \to 0$, yields $E\to -3D^2/7$.
The formal analogy between Eq.(69) and our expression Eq.(39), suggests
that the two results coincide provided
$B=\f{1}{2}\left[1+\f{E(\tau)}{D(\tau)^2}\right]$, which is indeed the case,
as noticed above.
The density skewness results
\be
S_{3\de}(\tau) = 4-2\,\f{E(\tau)}{D(\tau)^2}\;,
\ee
while the skewness of the velocity-divergence reads
(Bernardeau \etal 1994a):
\be
S_{3\eta}(\tau)= 2 - 2\,\f{E'(\tau)}{D(\tau)D'(\tau)}\;.
\ee
The divergence of the velocity field
$\theta$ in Bernardeau \etal (1994a) is defined in such a way that
$\theta = -f(\Om)\eta$ and thus $S_{3\theta}=-f(\Om)^{-1}S_{3\eta}$.\\

\subsubsection{Bernardeau (1994)}

The Eulerian perturbative expansion here is based on the spherical model; it
reads
\be
\de(\bfx,\tau) = D(\tau)\de_1(\bfx) + \f{1}{2!}D_2(\tau)\de_1(\bfx)^2
+\cdots \;,
\ee
explicitly up to the second-order contributions. The function
$D_2(\tau)$ is a solution of the differential equation
\be
\tau^2 D_2''+ \f{\tau(3+4\tau)}{2(1+\tau)}D_2'
-\f{3}{2(1+\tau)}D_2 = \f{3}{1+\tau}D^2+\f{8}{3}\tau^2
D'^{\,2}\;,
\ee
and, in the limit $\tau \to 0$, $D_2\to 34 D^2/21$.
The density skewness is
\be
S_{3\de}(\tau) = 3\,\f{D_2(\tau)}{D(\tau)^2}\;,
\ee
and the velocity-divergence one is
\be
S_{3\eta}(\tau)= -6+3\,\f{D'_2(\tau)}{D(\tau)D'(\tau)}\;.
\ee
Again, the divergence of the velocity field $\theta$ in Bernardeau (1994) is
such that $\theta = -f(\Om)\eta$. These Eulerian results for the skewness
are identical to those
obtained with Lagrangian methods by Bouchet \etal (1992): the expressions of
$S_{3\de}$ and $S_{3\eta}$, respectively in equations (71) and (72) and
equations (75)
and (76), coincide, in that the relation $D_2=\f{2}{3}(2D^2-E)$ actually
holds. \\

{}From this brief review we understand that, including ours, there are in
the literature four independent calculations of $S_{3\de}(\Om)$ and three of
$S_{3\eta}(\Om)$. In particular, comparing equation (59) with equations (67)
and (71), the two different expressions of $B(\tau)$ have been
used: {\it i)} $B_1(\tau)=\f{1}{4}\left[6-17\f{\kappa(\tau)}{D(\tau)}\right]$
is the equivalent of our function $B(\tau)$ in Martel \& Freudling (1991);
{\it ii)} $B_2(\tau)=\f{1}{2}\left[1+\f{E(\tau)}{D(\tau)^2}\right]$,
in Bouchet \etal (1992), coinciding with our $B(\tau)$, solution
of Eq.(46). The disagreement with the Martel \& Freudling results
is summarized in Fig. 1.

\begin{figure}
\vspace{6cm}

\caption{$\Om$-dependence of the density skewness $S_{3\delta}$.
The solid line represents our solution of eq.(59), which is equivalent to
that of Bouchet et al. (1992) and Bernardeau (1994). The dotted line shows
the corresponding expression obtained by Martel \& Freudling (1991).}
\end{figure}

It should however be noticed that, in the observationally relevant $\Om$
range, the two determinations of the skewness differ only by a few
percents, while the deviations become more conspicuous for low $\Om$ values.

\section{Summary and Conclusions}

In this paper we developed an Eulerian perturbative formalism which
allows to solve the Newtonian equations governing the dynamics of
a cold irrotational fluid in a generic Friedmann
universe, with vanishing cosmological constant. In particular,
the problem is notably simplified once described in terms of the
gravitational and velocity potentials, respectively $\varphi$ and $\Phi$.
The general set of scalar equations for these potentials was obtained and
it was shown that, in order to close this set one needs to
introduce an auxiliary scalar potential related to the current
$\bfj=\de\bfu$, carried by the density contrast $\de$, moving with
peculiar velocity $\bfu$.

Once the fundamental equations (22), which are nonlinear and nonlocal in
the potentials, were obtained,
the next step was to solve them for arbitrary density parameter $\Om$.
We solved them according to the general Eulerian
perturbative ansatz given in (24) and (25): apart from the
approximation intrinsic in any perturbative approach,
no further approximation was introduced. In particular, we focused in
seeking the second-order perturbative solutions $\varphi^{(2)}$
and $\Phi^{(2)}$,
whose explicit expressions are reported in equations (36) and (37),
which we demonstrated not to be separable with respect to their
space and time dependence;
the separability assumption is correct only in the case
of the Einstein-de Sitter universe, but not for more general
Friedmann models.

As a first application of our results, we have computed the second-order
deviation of the velocity-divergence, $\nabla\cdot\bfu$, from the density
contrast, $\de$. We finally calculated the skewness parameters $S_{3\de}$ and
$S_{3\eta}$ of the
density fluctuation and velocity-divergence fields, respectively: these
third-order moments quantify the asymmetric deformation of the original
Gaussian density and velocity-divergence distributions, induced by gravity
during the (weakly) nonlinear evolution. The weak dependence of
$S_{3\de}$ and $S_{3\eta}$ on the density parameter $\Om$,
first demonstrated applying Lagrangian perturbation theory, was confirmed.

Our results also clarify that the Eulerian perturbation theory, which is
characterized by non-separable perturbative higher-order solutions (in the case
of a general non flat Friedmann model), is equivalent to the corresponding
Lagrangian description, whose perturbative solutions of the dynamical
fluid equations are intrinsically separable with respect to the space
and time coordinates.

The present work provides a first attempt of building up a complete
higher-order Eulerian perturbation theory: most previous papers
only dealt with the construction of the density fluctuation and
velocity-divergence fields [a remarkable exception being the work by
Munshi \& Starobinsky (1994), which however, besides adopting a different
formalism is restricted to the flat case]. This kind of treatment should allow
a direct comparison of the Eulerian approach with the Lagrangian one, which
might lead to understanding many non-trivial aspects of the nonlinear
dynamics of self-gravitating collisionless matter in the expanding universe.

\section* {Acknowledgments} P.C. is grateful to George Efstathiou, James
Binney, Roman Juszkiewicz and Sergei Shandarin for stimulating discussions.
Thomas Buchert and Francis Bernardeau are warmly acknowledged for useful
comments and criticisms on a preliminary version of this paper.
This work has been partially supported by the
Fondazione Angelo Della Riccia and by
the EEC Human Capital and Mobility Programme. F.L., S.M. and L.M. acknowledge
the Italian MURST for partial financial support.

\appendix
\section{Approximate solutions of the Hamilton-Jacobi equation}

A standard method to deal with the solution of the Hamilton-Jacobi equation
(7) is through the simple ansatz
\be
\Phi_{\cal S} = {\cal W}(\bfx) - \calE D \;,
\ee
where ${\cal W}$ is Hamilton's principal function, and the constant
$\calE$ plays the role of particle `energy' per unit mass.
In our case, however, this
ansatz cannot work, since the peculiar gravitational potential,
self-consistently determined from the Poisson equation (3), generally
contains a non-trivial time dependence. Nevertheless, in most approximation
schemes aimed at moving particles, the physical peculiar gravitational
potential is practically replaced by some external, or `mock', potential,
e.g. derived from perturbation theory,
which allows to simply close the fluid dynamical equations, without fully
solving for the self-gravitation of the moving particles.
Examples of this are: {\it i)} the Zel'dovich approximation
(Zel'dovich 1970a,b), where one makes
the ansatz $\varphi(\bfx, D) = - \Phi(\bfx, D)$, and {\it ii)} the frozen flow
approximation (Matarrese et al. 1992), where one can write
$\varphi(\bfx, D) = \varphi^{(1)}(\bfx) - {D \over 3\,e(\Om)} \bigl(\nabla
\varphi^{(1)}(\bfx)\bigr)^2$ [extending to $\Om \neq 1$ the
expression in (Matarrese et al. 1992)].
Second-order type solutions can also be put in a similar form, but we will not
deal with this issue here.

Let us now analyze the two cases above in more detail.\\

\subsection{Zel'dovich approximation}

The action
in this case has to satisfy the standard Hamilton-Jacobi equation for
inertial motion,
\be
\f{\p\Phi_{\calS}}{\p D} + \f{1}{2}(\nabla\Phi_{\calS})^2 = 0\;,
\ee
which can be immediately solved in the form (A1) with
${\cal W}(\bfx) = - \varphi^{(1)}(\bfq) - \nabla_\bfq
\varphi^{(1)}({\bf q}) (\bfx - \bfq)$
and $\calE={1 \over 2}(\nabla_\bfq \varphi^{(1)}({\bf q}))^2$.
One then obtains the standard formula (e.g. Kofman 1991)
\be
\Phi_{\cal S}(\bfx,D)=-\varphi^{(1)}(\bfq) + {(\bfx - \bfq)^2 \over 2D}\;,
\ee
with $\bfx(\bfq,D)=\bfq-D\nabla_\bfq\varphi^{(1)}({\bf q})$.\\

\subsection{Frozen flow approximation}

In this case, after replacing the
appropriate expression for $\varphi$ in the Hamilton-Jacobi equation, we
immediately obtain
\be
\Phi_{\cal S}(\bfx,D)=-\varphi^{(1)} (\bfx)\;,
\ee
corresponding to vanishing particle `energy', $\calE=0$. This conclusion is
consistent with a simple picture of this approximation according to which
the fluid elements behave as test particles, which just trace the initial
stream-lines.

The advantage of formulating these two approximations in terms of the
Hamilton-Jacobi approach is that, once the solution is known, both mass and
momentum conservation are exactly satisfied at all stages of the
nonlinear evolution.

\section{Failure of the separability ansatz in non-flat models}

In this appendix we explicitly demonstrate that, in the case
of a generic Friedmann universe, any separable perturbative
approximation is $not\,$ a solution of the dynamical equations for the
cosmological potentials $\varphi$ and $\Phi$. To avoid an exceeding
proliferation of mathematical symbols, we use in this appendix the same
notation of the main text.

Let us therefore suppose that a possible class of solutions of the
fundamental equations for the cosmological potentials may be
written in the form
\be
\varphi(\bfx,D)=\varphi^{(1)}(\bfx) +
\varphi^{(2)}(\bfx,D) +
\cdots\equiv\varphi_1(\bfx)+D\,g_\varphi(D)\,\varphi_2(\bfx) + \cdots\;,
\ee
\be
\Phi(\bfx,D)=
\Phi^{(1)}(\bfx)+\Phi^{(2)}(\bfx,D)+\cdots
\equiv\Phi_1(\bfx)+D\,g_\Phi(D)\,\Phi_2(\bfx)+\cdots\;,
\ee
where the functions $\varphi_2(\bfx)$ and $\Phi_2(\bfx)$ are assumed to be
{\it independent} of $D$, while the functions $g_\varphi(D)$ and $g_\Phi(D)$
are assumed to depend only on $D$. The calculations performed in Section 2 can
then be repeated up to Eqs.(36) and (37), which we rewrite in the form
\be
g_\varphi(D)\varphi_2(\bfx)= - \Big[1-B(D)\Big]F_2(\bfx)+
\f{1}{2}B(D)\Big[\nabla\varphi_1(\bfx)\Big]^2\;
\ee
and
\be
g_\Phi(D)\Phi_2(\bfx) = \Big[1-2B(D)-DB'(D)\Big]F_2(\bfx)-
\f{1}{2}\Big[2B(D)+DB'(D)\Big]\Big[\nabla\varphi_1(\bfx)\Big]^2\;,
\ee
where the two space-dependent functions $\varphi_2(\bfx)$ and $\Phi_2(\bfx)$
can be easily determined from the $\Om \to 1$ limit (see Eqs.(A14) and (A15)
below). It is then clear that the two functions $g_\varphi(D)$ and
$g_\Phi(D)$ cannot be
space-independent, as they should, unless, either one restricts to the
Einstein-de Sitter case,in which case $g_\varphi(D)\equiv g_\Phi(D)
\equiv 1$, or one imposes suitable restrictions on the functional form of the
initial potential $\varphi_1(\bfx)$. Specifically, it is
easy to show that the separability ansatz (A5), (A6) for the
gravitational and velocity potentials requires the validity of the
constraint
\be F_2(\bfx)= K \Big[\nabla\varphi_1(\bfx)\Big]^2\;,
\ee
with $K$ a suitable constant.
It is interesting to note that a sufficient condition for the validity of this
constraint is that the second-order invariant of the initial deformation
tensor vanishes, $\mu_2(\bfu_1)=0$ (this is e.g. the case of
one-dimensional perturbations, which are described by the Zel'dovich
solution). This can be easily shown by replacing the relation (A9) into
the last of Eqs.(26).

\section{Second-order Eulerian perturbation theory in an
Einstein-de Sitter model}

In this appendix, we want to apply our formalism to the case of an
Einstein-de Sitter universe, without cosmological constant. In particular, and
as an example of how our formalism works, we
want to derive the expressions of $\varphi^{(2)}(\bfx, a)$ and
$\Phi^{(2)}(\bfx,a)$, then the second-order corrections
$\de^{(2)}(\bfx,a)\equiv a^2\,\de_2(\bfx)$ and
$\bfu^{(2)}(\bfx,a)\equiv a\,\bfu_2(\bfx)$ already known in the literature.
Again, we use in this appendix the same notation of the main text.

The fundamental equations for the cosmological potentials $\varphi$ and $\Phi$
are given by
\be
\left\{
\begin{array}{l}
{\displaystyle \f{\p\Phi}{\p a}}+
{\displaystyle \f{1}{2}}(\nabla\Phi)^2 +
{\displaystyle \f{3}{2a}}(\Phi + \varphi)=0\;, \\ \\
{\displaystyle \f{\p}{\p a}}(a\varphi)+\Phi+F=0\;, \\ \\
\nabla^2F=a\nabla\cdot\left[\nabla^2\varphi\nabla\Phi\right]\;.
\end{array}
\right.
\ee

As in Section 3, to find approximate perturbative solutions of these equations,
we expand $\varphi$, $\Phi$ and $F$ according to the relations
\be
\varphi(\bfx,a) = \varphi^{(1)}(\bfx) +\varphi^{(2)}(\bfx,a)+\cdots \equiv
\varphi_1(\bfx) + a\,\varphi_2(\bfx) +\cdots\;,
\ee
\be
\Phi(\bfx,a) = \Phi^{(1)}(\bfx) +\Phi^{(2)}(\bfx,a)+\cdots \equiv
\Phi_1(\bfx) + a\,\Phi_2(\bfx) +\cdots\;,
\ee
and
$F(\bfx,a) = F^{(2)}(\bfx,a) \equiv a\,F_2(\bfx)$;
we remind that
$\de^{(1)}(\bfx, a)=a\nabla^2\varphi^{(1)}(\bfx)$, while
$\de_1(\bfx)=\nabla\varphi_1(\bfx)$,
where $\de^{(1)}(\bfx, a)=a\,\de_1(\bfx)$;
$\varphi_1(\bfx)=-\Phi_1(\bfx)$. Remind that $F$ is by definition
at least a second-order quantity. Finally, we stress that the general
expansions (24) and (25) have become separable in space and time in
the flat case. In terms of the perturbed quantities one has
\be
\left\{
\begin{array}{l}
{\displaystyle \f{5}{2}}\Phi_2+
{\displaystyle \f{1}{2}}(\nabla\Phi_1)^2 +
{\displaystyle \f{3}{2}}\varphi_2 = 0\;, \\ \\
2\varphi_2 +\Phi_2 +F_2 = 0\;, \\ \\
\nabla^2F_2 = -\de_1^{\,2}+(\bfu_1\cdot\nabla)\,\de_1\;.
\end{array}
\right.
\ee
{}From the first two equations it is possible to obtain the expressions for
$\varphi_2$ and $\Phi_2$:
\be
\varphi_2 = -\f{5}{7}\,F_2 + \f{1}{7}\,(\nabla\varphi_1)^2\;,
\ee
\be
\Phi_2 = \f{3}{7}\,F_2 - \f{2}{7}\,(\nabla\varphi_1)^2\;.
\ee
We stress again that the velocity field $\bfu$ is manifestly irrotational
in physical space, namely
\be
\bfu(\bfx, a) = \nabla\left[\Phi^{(1)}(\bfx) + \f{3}{7}\, F^{(2)}(\bfx, a)
-\f{2}{7}\,a\,\bfu^{(1)}(\bfx)^2\right]\,.
\ee
The final expressions for $\de$ and $\nabla\cdot\bfu$ are
\begin{eqnarray}
\de
& = & \de^{(1)}+
      \nabla^2\left[-\f{5}{7}\,a\,F^{(2)}+\f{1}{7}\,a^2\,\bfu^{(1)\,2}\right]
\nonumber \\
& = & \de^{(1)}(\bfx, a) +
\f{5}{7}\,\de^{(1)}(\bfx, a)^2-a\,(\bfu^{(1)}\cdot\nabla)\,\de^{(1)}(\bfx, a) +
\f{2}{7}\,a^2\,\sum_{\al\beta}\Bigl(\p_{\beta}\,u^{(1)}_{\al}\Bigr)^2\;,
\end{eqnarray}
and
\begin{eqnarray}
-a\nabla\cdot\bfu
& = & \de^{(1)}+
      \nabla^2\left[-\f{3}{7}\,a\,F^{(2)}+\f{2}{7}\,a^2\,\bfu^{(1)\,2}\right]
\nonumber \\
& = & \de^{(1)}(\bfx, a) +
\f{3}{7}\,\de^{(1)}(\bfx, a)^2-a\,(\bfu^{(1)}\cdot\nabla)\,\de^{(1)}(\bfx, a) +
\f{4}{7}\,a^2\,\sum_{\al\beta}\Bigl(\p_{\beta}\,u^{(1)}_{\al}\Bigr)^2\;,
\end{eqnarray}
which give equations (42) and (43) of the main text; they can
be also compared with equation (18.8) in Peebles (1980), where, however, a
slightly different
definition of gravitational potential is used, with equations (19a,b) in Wise
(1988) and with equations (8) and (10) in Catelan \& Moscardini (1994b).

\section{Second-order Eulerian perturbation theory in a closed
Friedmann model}

The $\Om > 1$ case may be treated starting from
equation (44) and transforming to the new variable (see e.g. Peebles 1980)
\be
\tau \equiv  1-\f{1}{\Om}\;,
\ee
where now $0\leq \tau < 1$, the lower limit corresponding to the
Big Bang, the upper limit to the instant of
maximum expansion: in this appendix we consider only the expanding phase.

The equation to be integrated is formally identical to equation (46) and the
initial
conditions are the same. Finally, all the non-trivial expressions valid for a
closed model are
\be
J(\tau) \equiv \f{D(\tau)}{D'(\tau)} = \tau\,f(\tau)^{-1}\;,
\ee
and $J(\tau) \to \tau$ in the limit $\tau \to 0$, like
in the open models; furthermore
\be
D(\tau)= -1 + \f{3}{\tau} - \f{3\sqrt{1-\tau}}{\tau^{3/2}}\,M(\tau)\;,
\ee
with $M(\tau)\equiv{\rm arctan}\sqrt{\tau/(1-\tau)}\,$;
$D(\tau)=2\tau/5$ in the limit $\tau\to 0$, as in the case $\Om<1$.
The function $A(\tau)$ is now given by
\be
A(\tau)=2+\f{3}{2}\Big[(1-\tau)\,f(\tau)^2\Big]^{-1}\;,
\ee
and
\be
f(\tau) = -\f{3}{2}\left[
\f{3\sqrt{\tau(1-\tau)} - (3-2\tau)M(\tau)}
{(3-\tau)\sqrt{\tau(1-\tau)}-(3-3\tau)M(\tau)}
\right]\;,
\ee
with $f(0) = 1$. We then have:
\be
B(\tau) = \f{1}{2} -\f{1}{4D(\tau)^2}+\f{9}{4\tau D(\tau)^2}
\left\{1-\sqrt{\f{1-\tau}{\tau}}\,M(\tau)-\f{1}{2}
\left[\sqrt{\f{1-\tau}{\tau}}-\f{M(\tau)}{\tau}
\right]^2
\right\}\;.
\ee
If, alternatively, one prefers to work directly in terms of
the time-variable $\Om$, the formulae which have to be used are, respectively:
\be
J(\Om)\equiv \f{D(\Om)}{D'(\Om)} = \Om(\Om-1)\,f(\Om)^{-1}\;,
\ee
\be
D(\Om) = \f{1}{(\Om-1)^{3/2}}\left[(1+2\Om)\sqrt{\Om-1}-3\Om\, M(\Om)
\right]\;,
\ee
\be
f(\Om) = -\f{3\Om}{2}
\left[\f{3\sqrt{\Om-1}-(2+\Om)\,M(\Om)}{(1+2\Om)\sqrt{\Om-1}-3\Om\, M(\Om)}
\right]\;,
\ee
where $M(\Om)\equiv{\rm arctan}\sqrt{\Om-1}$ and $M(1)=0$.

\end{document}